\documentclass[authoryear,preprint,review,12pt]{elsarticle}

\usepackage{graphics}

\usepackage{graphicx}

\usepackage{epsfig}

\usepackage{amssymb}

\usepackage{amsthm}

\usepackage{lineno}

\journal{New Astronomy}

\begin{document}

\begin{frontmatter}



\title{A Clue To Magnetic Activity Occurring \\ On Near-Contact Binary V1003\,Her}


\author{H.A. Dal\corref{cor1}}
\ead{ali.dal@ege.edu.tr}
\cortext[cor1]{Corresponding author}

\author{E. Sipahi}

\address{Ege University, Science Faculty, Department of Astronomy and Space Sciences, 35100 Bornova, \.{I}zmir, Turkey}

\begin{abstract}

Taking into account results obtained from models and analyses, we determined the nature and structures of V1003\,Her. We analysed the light curves obtained in this study together with the available radial velocity curve. The analysis revealed that the inclination ($i$) of the system was found to be 45$^{\circ}$.82$\pm$2$^{\circ}$.71, while the semi-major axis ($a$) was computed as 2.925$\pm$0.143 $R_{\odot}$. The mass of the primary component was found to be 1.204$\pm$0.004 $M_{\odot}$, while it was obtained as 0.450$\pm$0.006 $M_{\odot}$ for the secondary component. The radius of the primary component was computed as 1.424$\pm$0.001 $R_{\odot}$, while it was computed as 0.904$\pm$0.001 $R_{\odot}$ for the secondary component. The analysis indicates that there are two spotted areas close to each other on the surface of the secondary component. In addition, B-band light and $B-V$ colour curves reveal that there is a radiation excess toward the phase of 0.40. This should be caused due to some heating of the surface by the events like thermal flare, or the structures like plage. Finally, the analysis and model demonstrated that V1003\,Her should be a near-contact binary with high level magnetic activity on its surface.

\end{abstract}

\begin{keyword}

techniques: photometric --- (stars:) binaries: eclipsing --- stars: late-type --- stars: individual: (V1003 Her)

\end{keyword}

\end{frontmatter}



\section{Introduction}
V1003\,Her (=GSC\,01596-01204) is classified as a variable star of $\delta$ Scuti type in the SIMBAD Database. The variability seen in the system was found by Hipparcos satellite for the first time. V1003\,Her was classified as a contact binary with a period of 0$^{d}$.4933 from the spectral type A7 by \citet{Due97}. Then, \citet{Hog00} listed the system in the TYCHO-2 Catalogue, in which the brightness and colour are given as $V=9^{m}.79$ and $(B-V)=0^{m}.41$. However, citing \citet{Bre94, Bre95}, \citet{Rod00} listed V1003\,Her as a $\delta$ Scuti star with a period of 0$^{d}$.2467. The system was observed in both the ASAS-3 Survey \citep{Poj97, Poj02} and the Northern Sky Variability Survey (hereafter NSVS) \citep{Woz04}. V1003\,Her was listed as ASAS\,185318+2113.5 in the ASAS-3 database, while as NSVS\,11074663 in the NSVS database. The data obtained in the ASAS-3 indicated the most probable period is 21$^{d}$.846. \citet{Ruc08} obtained the radial velocity curves of the system with $K_{1}=64.07\pm0.94$ $kms^{-1}$, $K_{2}=171.91\pm0.94$ $kms^{-1}$ and $\nu_{0}=-6.92\pm0.61$ $kms^{-1}$. They gave the light elements of the system as $HJD_{min}=2448500.43+0^{d}.493322$, while they mentioned that the system is from the spectral type of A7, indicating that this spectral type is in agreement with the colours given in the 2MASS Catalogue \citep{Cut03}. \citet{Deb11} tried to analysis the data given in the ASAS-3 database. They found 3 possible periods such as $0^{d}.246657$, $0^{d}.395712$ and $0^{d}.493326$. Then, they tried to analysis the mono-colour light curve obtained from the ASAS-3 database, which was phased with the period of $0^{d}.493326$, and they classified the system as a B-type contact binary. However, appraising both ASAS-3 and NSVS data, \citet{Ruc08} commentated that a detailed photometric observations are needed.

In this study, we obtained more detailed multi-colour light curves of the system. We adjusted the light elements using available minima times. Then, analysed the multi-colour light curves together with the radial velocity obtained by \citet{Ruc08} simultaneously. Our more detailed observations indicated that V1003\,Her is a near-contact binary with high level magnetic activity on its surface.

\section{Observations}
Observations of the system were acquired with a thermoelectrically cooled ALTA $U+47$ $2048\times2048$ pixel CCD camera attached to a 40 cm - Schmidt - Cassegrains - type MEADE telescope at Ege University Observatory. The observations were continued in BVR bands during four nights in 04, 18, 21 and 29 July 2011. Some basic parameters of program stars are listed in Table 1. The names of the stars are listed in first column, while J2000 coordinates are listed in next two columns. The V magnitudes are in fourth column, and B-V colours are listed in the last column. In table, the coordinates, brightness and colours were taken from the SIMBAD database.

Although the program and comparison stars are very close on the sky, differential atmospheric extinction corrections were applied. The atmospheric extinction coefficients were obtained from observations of the comparison stars on each night. Heliocentric corrections were also applied to the times of the observations. The mean averages of the standard deviations are 0$^{m}$.023, 0$^{m}$.011, and 0$^{m}$.010 for observations acquired in the BVR bands, respectively. To compute the standard deviations of observations, we used the standard deviations of the reduced differential magnitudes in the sense comparisons (HD\,343345) minus check (HD\,343344) stars for each night. There was no variation observed in the standard brightness comparison stars.

There are a few light curves, such as those obtained from the ASAS-3 and the NSVS database, and so there are two trustable minima times in the literature, which were given by \citet{Ruc08} and \citet{Deb11}. In this study, we obtained new four minima times.

\begin{center}
\begin{equation}
Min~I~(Hel.)~=~24~55761.3729(5)~+~0^{d}.493321(1)~\times~E
\end{equation}
\end{center}

Using these six minima times listed in Table 2, we adjusted the light elements as given in Equation (1). Using these light elements, all observations were phased. The obtained light and colour curves are shown in Figure 1. In the figure, B light curve is seen in the upper panel, while the colour curves are shown in the bottom panel. In the bottom panel of the figure, $B-V$ curve is upper one, while the $V-R$ curve is bottom one.

\section{Light Curve Analysis}
The light curves obtained from both ASAS-3 data and NSVS data are rather noisy to reveal the general properties of the system. In fact, \citet{Deb11} did not give any information about system apart from being a B-type contact binary. As seen from Figure 1, there is a noticeable distortion from the phase of 0.20 to 0.40.  To understand the reason(s) of this distortion, we tried to model the light curves obtained in this study with the Wilson-Devinney Code \citep{Wil71, Wil90}. For this aim, we analysed the light curves obtained in the BVR bands together with the available radial velocity curve simultaneously, using the PHOEBE V.0.31a software \citep{Prs05}, which is used the version 2003 of Wilson-Devinney Code \citep{Wil71, Wil90}. We tried to analyse the light curves in several different modes, such as the "overcontact binary not in thermal contact" and "double contact binary" modes. The initial analyses demonstrated that an astrophysical acceptable result can be obtained if the analysis is carried out in the "overcontact binary not in thermal contact" mode, while no acceptable results could be obtained in the other mode.

\citet{Amm06} determined the temperature of the system as 6252 K. Thus, the temperature of the primary component was fixed to 6252 K in the analyses, and the temperature of the secondary was taken as a free parameter. Considering the spectral type corresponding to this temperature, the albedos ($A_{1}$ and $A_{2}$) and the gravity darkening coefficients ($g_{1}$ and $g_{2}$) of the components were adopted for the stars with the convective envelopes \citep{Luc67, Ruc69}. The non-linear limb-darkening coefficients ($x_{1}$ and $x_{2}$) of the components were taken from \citet{Van93}. In the analyses, the fractional luminosity ($L_{1}$) of the primary component, the inclination ($i$), mass ratio ($q$), semi-major axis ($a$) were also taken as the adjustable free parameters.

The inclination ($i$) of the system was found to be $45^\circ.819\pm2^\circ.712$, while the mass ratio of the system was found to be 0.374$\pm$0.008, and the semi-major axis ($a$) was computed as 2.925$\pm$0.143 $R_{\odot}$. In Table 3, all the parameters derived from the analyses are listed in detail, while the synthetic light curves are shown in Figures 2 and 3. In addition, we also derived the 3D model of Roche geometry, using the parameters obtained from the light curve analysis. The derived 3D model of Roche geometry is shown in Figure 4.

Depending on its surface temperature, the mass of the primary component was found to be 1.204$\pm$0.004 $M_{\odot}$, while the mass of the secondary component was found to be 0.450$\pm$0.006 $M_{\odot}$. Considering the semi-major axis, the radius of the primary component was computed as 1.424$\pm$0.001 $R_{\odot}$, while it was computed as 0.904$\pm$0.001 $R_{\odot}$ for the secondary component. In addition, the luminosity of the primary component was computed as 2.783$\pm$0.007 $L_{\odot}$, and it was computed as 1.078$\pm$0.004 $L_{\odot}$ for the secondary component.

We compared the absolute properties of the system with the others in some planes, such as the mass-radius ($M-R$), mass-luminosity ($M-L$), and luminosity-effective temperature ($L-T_{eff}$) planes. All the comparisons are shown in Figure 5. In the figures, the lines represent the ZAMS theoretical model developed for the stars with $Z=0.02$ by \citet{Gir00}, while dashed lines represent the TAMS theoretical model. The filled circles represent the primary components, while the open circles represents the secondary ones. The components of V1003\,Her are located together with some samples of its analogues, such as YY\,CrB, DN\,Boo, CK\,Boo, $\epsilon$ CrA, FG\,Hya, TV\,Mus, AW\,UMa, GR\,Vir, V776\,Cas.  The sample systems were taken from \citet{Ess10}, and they are shown in purple colour, while V1003\,Her is shown in black colour in Figure 5.

\section{Discussion}
The analyses of the detailed observations indicated that V1003 Her is very close to the edge of eclipse due to its the inclination ($i$) of $45^\circ.819\pm2^\circ.712$ in these radii. According to the results, the system is a chromospherically active binary system, though the components are the evolved stars.

Using all available minima times, the orbital period was adjusted as $0^{d}.493321$, and all the data obtained in this study were phased. We analysed the multi-band light curve simultaneously with available radial velocity curve obtained by \citet{Ruc08}. The initial attempt of the analysis revealed that an acceptable results could be obtained in the case of analysis carrying out in the overcontact binary not in thermal contact mode. The mass ratio was found to be 0.374$\pm$0.008 and the semi-major axis ($a$) was found to be 2.925$\pm$0.143 $R_{\odot}$. Assuming that the temperature of the primary component is 6252 K as determined by \citet{Amm06}, the temperature of the secondary component was found to be 6192$\pm$144 K. In this case, the fractional radii were found to be $r_{1}=0.487\pm0.001$ for the primary component and $r_{2}=0.309\pm0.001$ for the secondary one. Here, the sum of fractional radii is found to be so close to $r_{1}+r_{2}=0.80$. In this point, according to \citet{Kop56}'s criteria for the overcontact systems, V1003\,Her seems to be a very strong candidate for the W\,UMa type binaries. Computing the absolute parameters of V1003\,Her, we compared the system with its analogues in some planes, such as the mass-radius ($M-R$), mass-luminosity ($M-L$), and luminosity-effective temperature ($L-T_{eff}$). As it is seen from Figure 5, the components of the system are in agreement with their analogues. In all panels, V1003\,Her seems to be a bit an evolved system.

However, as it is clearly seen from Figures 1, there is a dramatic asymmetry in the light curves. Because of this, we assumed that the secondary component has two cool spots on its surface to remove this asymmetry. The surface temperatures of the components are so close to each other, and so both of them can be a magnetically active star. However, it is well known that the spot solution suffers from non-uniqueness, we just assumed that the active one is the secondary star. In addition, assuming spotted star is the primary one, a similar solution can possibly be obtained.

On the other hand, as it is seen from Figure 3, the synthetic light curves do not absolutely fitted the observations between the phases of 0.20 and 0.40 for the B and V-bands. The synthetic curves located below the observations in the phase range. The case indicated that there is a radiation excess between this phase range. In fact, as it is seen from both $B-V$ and $V-R$ colour curves shown in Figure 1, the colours are also getting bluer towards this phase range. In addition, the $B-V$ colour is getting much bluer than the $V-R$ colour. Both the radiation excess in the light curves and the blue excess in the colours indicate any hot area on the surface of one of the component. Considering the existence of cool spots, this heating area should be related and caused by the magnetically activity. The case is a general phenomenon for the stars like this system. Chromospherically active stars, whose ages are close to solar age or over, exhibit some hot spots, such as faculae on the photosphere and plages in the chromosphere, as well as cool spots on the photosphere. Some of them also exhibit even the large flares \citep{Wil94, Ber05}. There are several analogues of these stars, such as X\,Tri \citep{Str09}, $\lambda$ And, II\,Peg \citep{Fra08}, V471\,Tau \citep{Iba05}.

Although we tried to modelled the blue excess seen in the light curves with some hot spots as some structures like faculae, the synthetic curves obtained from these models do not simultaneously fit all the BVR observations in one model. In the analyses, although some attempts could give a solution for the V or R-bands, we could not obtained any acceptable solution for B-band in one model derived in the PHOEBE V.0.31a software. The opposite cases were also seen in some attempts. The case indicates that the blue excess is not caused by any hot spot as some structures like faculae. These distortions in the light curves should be caused by any slow flare \citep{Dal10} rather than any structures like faculae. At the moment, there is no enough data to be sure that V1003\,Her exhibits any flare like activity. However, a similar case was obtained in the case of the totally-eclipsing binary, GSC\,4589-2999, by \citet{Sip13}. The same colour and light variations have been detected in the observations of GSC\,4589-2999, though geometrical configurations of two systems are remarkable different from each other. Considering the results obtained from the PHOEBE analyses, it is most likely that the variation seen in the case of V1003\,Her is caused by a flare like event occurring on the surface of the active component. According to \citet{Dal10}, the slow flares can cause some effects on both light and colour variations. However, it needs to much more photometric observations to be sure.

\section*{Acknowledgments} The author acknowledges the generous observing time awarded to the Ege University Observatory. We also thank the referee for useful comments that have contributed to the improvement of the paper.

\clearpage

\begin{figure*}
\hspace{1.0 cm}
\includegraphics[width=14cm]{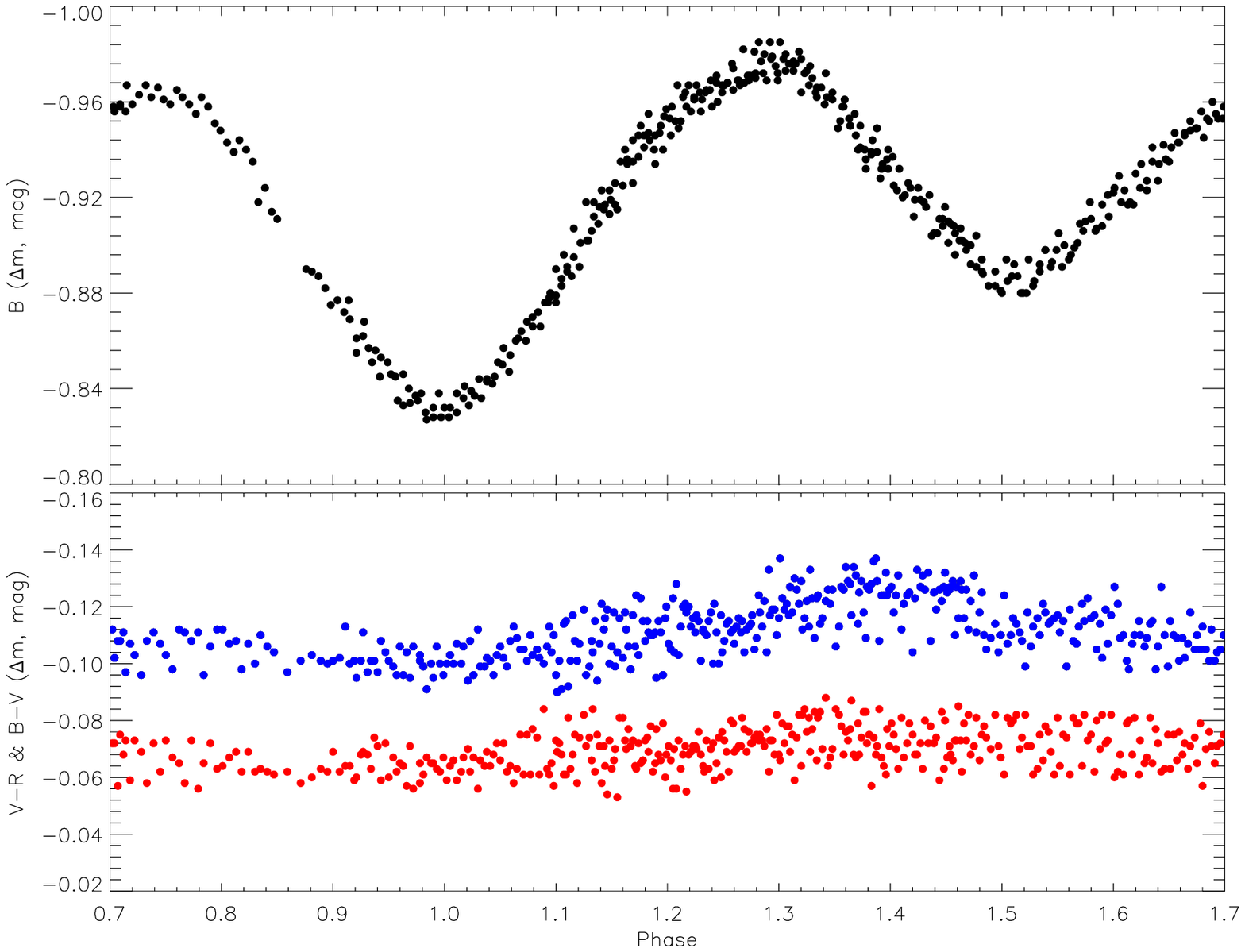}
\vspace{0.2 cm}
\caption{The B light, $B-V$ and $V-R$ colour curves of V1003\,Her obtained at Ege University Observatory. In the upper panel, the filled circles represent B-band observations. In the bottom panel, the filled-blue circles (upper curve) represent $B-V$ colour and filled-red circles (bottom curve) represent $V-R$ colour variations, respectively.}
\label{Fig.1}
\end{figure*}

\begin{figure*}
\hspace{1.3 cm}
\includegraphics[width=15cm]{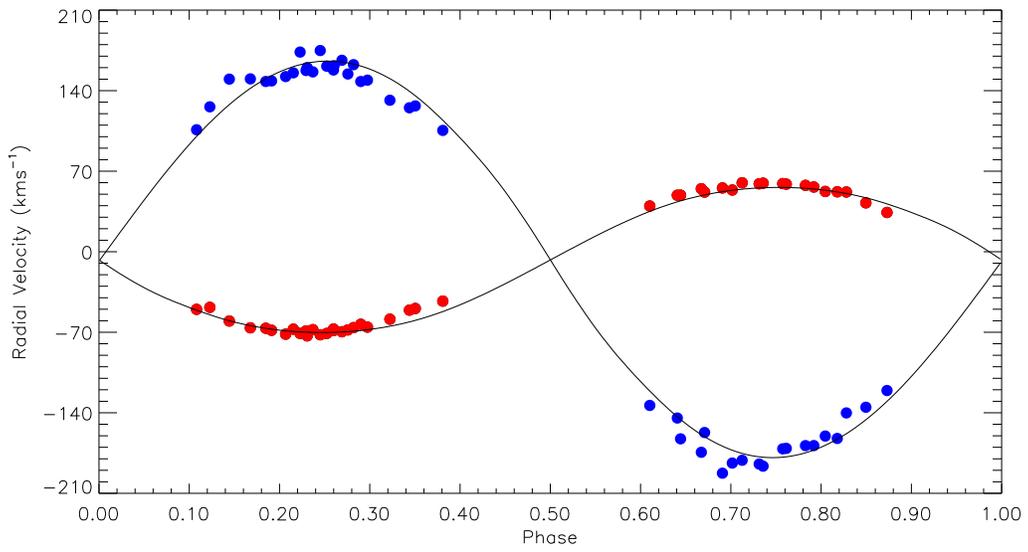}
\vspace{0.2 cm}
\caption{Radial velocity curve of V1003\,Her. Filled circles represent the observations of the primary, while open circles represent the secondary components. Solid curves are the theoretic radial velocity curves derived by the light curve analysis.}
\label{Fig.2}
\end{figure*}

\begin{figure*}
\hspace{2.5 cm}
\includegraphics[width=19cm]{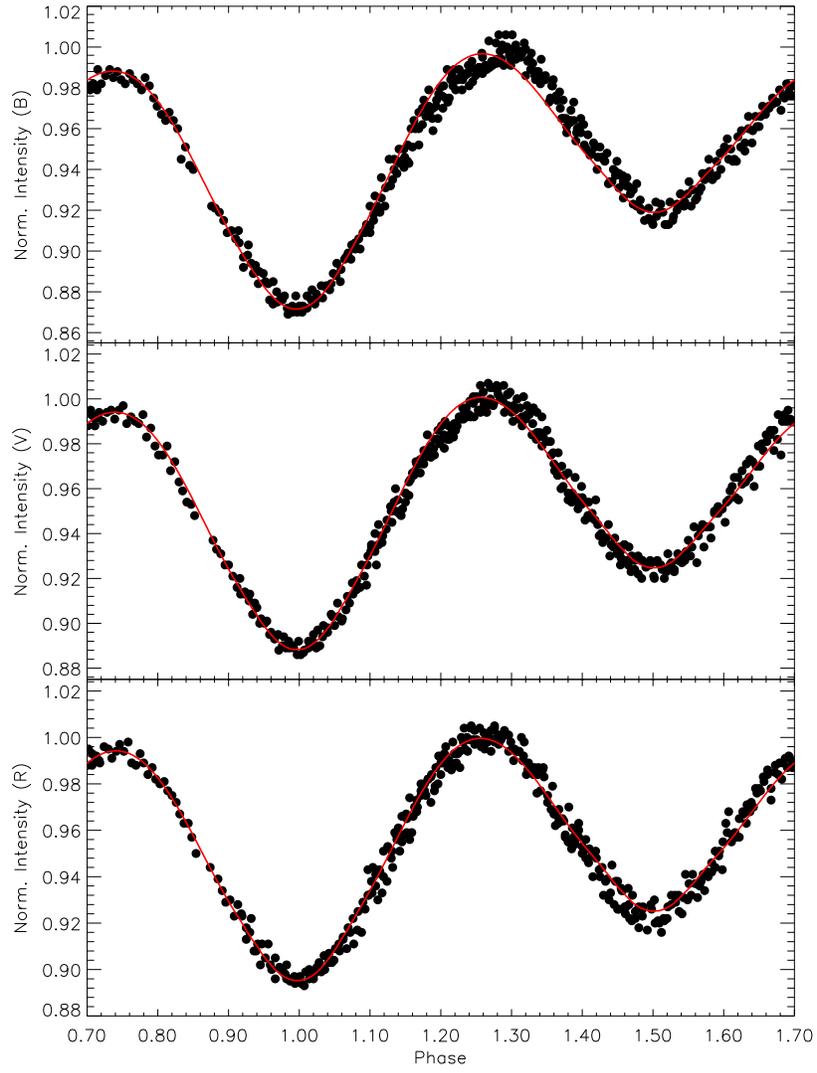}
\vspace{0.2 cm}
\caption{The BVR light curves of V1003\,Her and the synthetic solutions for the observations in each band.}
\label{Fig.3}
\end{figure*}

\begin{figure*}
\hspace{2.2 cm}
\includegraphics[width=9cm]{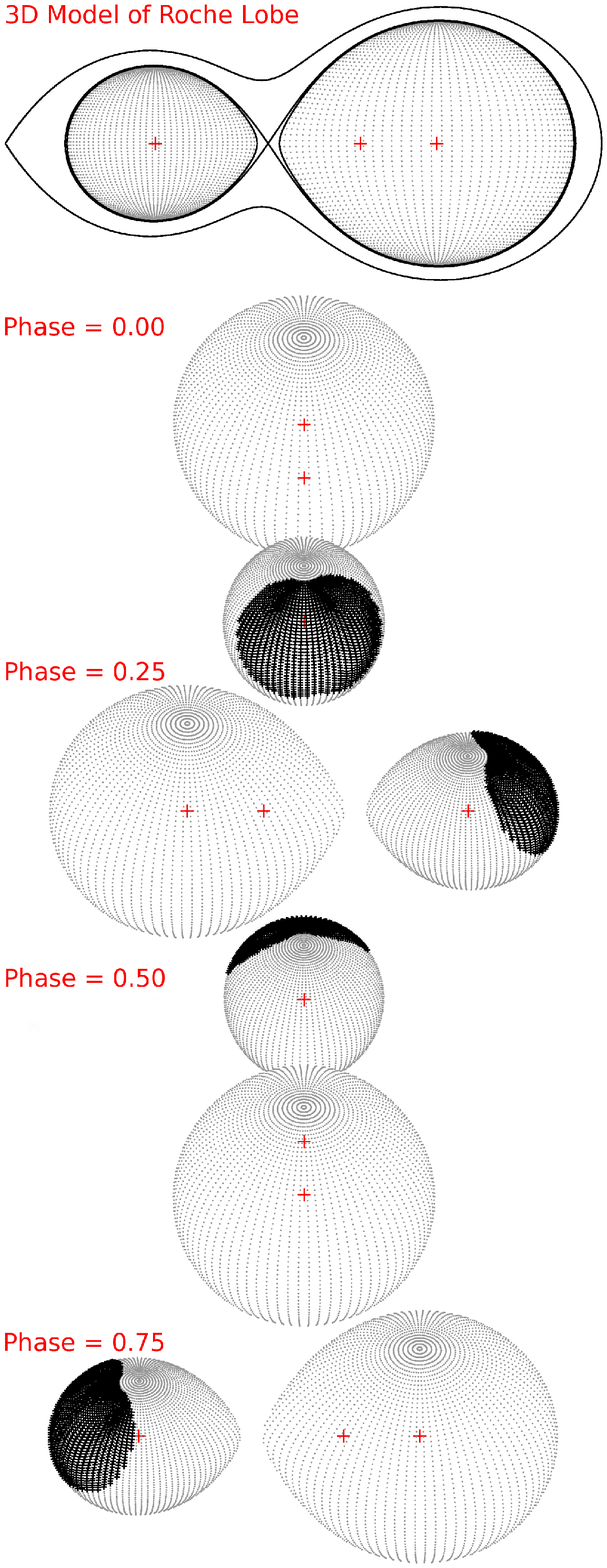}
\vspace{0.2 cm}
\caption{The 3D model of Roche geometry for V1003 Her.}
\label{Fig.4}
\end{figure*}

\begin{figure*}
\hspace{2.5 cm}
\includegraphics[width=20cm]{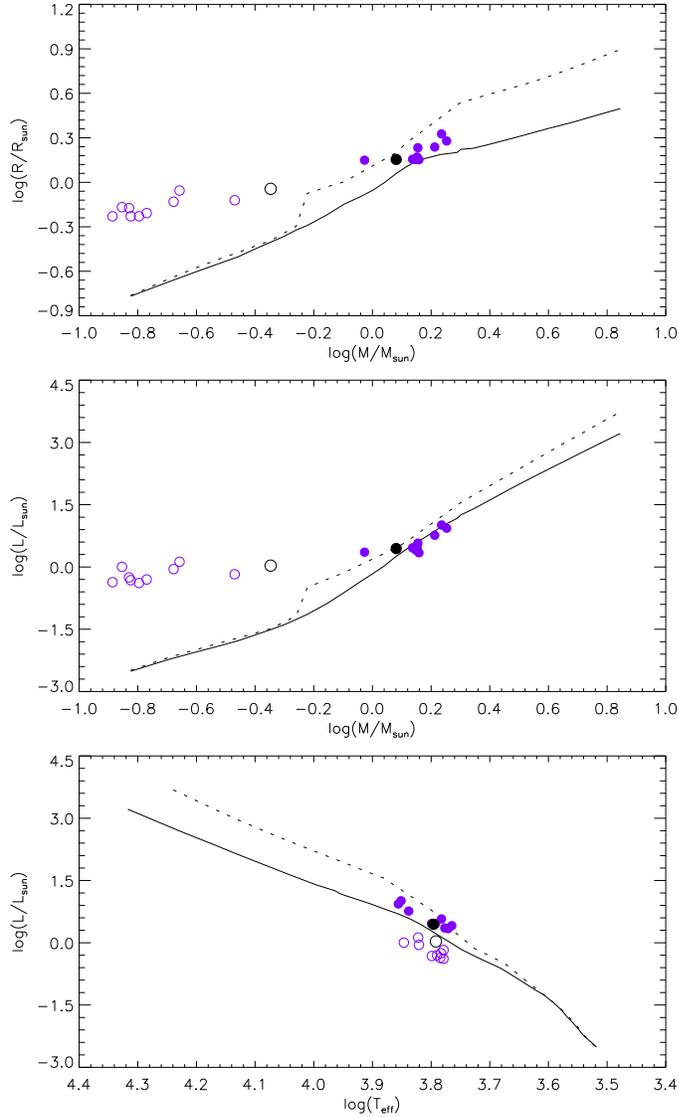}
\vspace{0.2 cm}
\caption{The places of the components of V1003 Her in the planes of (a) the mass-radius ($M-R$), (b) mass-luminosity ($M-L$), and (c) luminosity-effective temperature ($L-T_{eff}$). In the panels, the continuous and dashed lines represent the ZAMS and TAMS theoretical models developed by \citet{Gir00}, respectively. The filled circles represent the primary components, while the open circles represent the secondary ones. The dark circles represent the V1003\,Her components, while the purple coloured circles represent the components of other contact binaries.}
\label{Fig.5}
\end{figure*}

\clearpage

\begin{table*}
\centering
\caption{Basic parameters for the observed stars.}
\vspace{0.3cm}
\begin{tabular}{lcccc}
\hline\hline
Star	&	Alpha (J2000)	&	Delta (J2000)	&	V	&	$B-V$	\\
Name	&	($^{h}$ $^{m}$ $^{s}$)	&	($^{\circ}$ $^{\prime}$ $^{\prime\prime}$)	&	(mag)	&	(mag)	\\
\hline									
V1003\,Her	&	18 53 17.543 	&	+21 13 32.74	&	9.79	&	0.41	\\
HD\,343345	&	18 54 08.901 	&	+21 12 38.14	&	10.61	&	0.57	\\
HD\,343344	&	18 54 12.639	&	+21 16 12.48	&	10.09	&	0.63	\\
\hline
\end{tabular} 
\end{table*}

\begin{table*}
\centering
\caption{The minima times and ($O-C$) residuals. In the first column, the standard deviations of obtained minima times are given in the brackets near themselves.}
\vspace{0.3cm}
\begin{tabular}{lrcccr}
\hline\hline

O	&	E	&	$(O-C)_{II}$	&	Type	&	Method	&	REF	\\
\hline											
48500.4300 (-)	&	-14718.47	&	0.0163	&	II	&	-	&	Rucinski et al. (2008)	\\
53099.9063 (-)	&	-5394.99	&	0.0049	&	I	&	-	&	Deb and Singh (2011)	\\
55747.3143 (2)	&	-28.50	&	0.0004	&	II	&	BVR	&	This Study	\\
55761.3736 (1)	&	0.00	&	0.0000	&	I	&	BVR	&	This Study	\\
55764.3325 (1)	&	6.00	&	-0.0010	&	I	&	BVR	&	This Study	\\
55772.4711 (1)	&	22.50	&	-0.0022	&	II	&	BVR	&	This Study	\\
\hline
\end{tabular} 
\end{table*}

\begin{table*}
\centering
\caption{The parameters of components obtained from the light curve analysis.}
\vspace{0.3cm}
\begin{tabular}{lrr}
\hline\hline

Parameter	&	Value	\\
\hline			
$q$	&	0.374$\pm$0.008	\\
$a$ ($R_{\odot}$)	&	2.925$\pm$0.143	\\
$i$ ($^\circ$)	&	45.819$\pm$2.712	\\
$T_{1}$ (K)	&	6252 (Fixed)	\\
$T_{2}$ (K)	&	6192$\pm$144	\\
$\Omega_{1}$	&	2.642$\pm$0.041	\\
$\Omega_{2}$	&	2.642 (Fixed)	\\
$L_{1}/L_{T}$ (B)	&	0.726$\pm$0.006	\\
$L_{1}/L_{T}$ (V)	&	0.723$\pm$0.004	\\
$L_{1}/L_{T}$ (R)	&	0.722$\pm$0.005	\\
$g_{1}$, $g_{2}$	&	0.32, 0.32 (Fixed)	\\
$A_{1}$, $A_{2}$	&	0.50, 0.50 (Fixed)	\\
$x_{1,bol}$, $x_{2,bol}$	&	0.644, 0.644 (Fixed)	\\
$x_{1,B}$, $x_{2,B}$	&	0.816, 0.816 (Fixed)	\\
$x_{1,V}$, $x_{2,V}$	&	0.726, 0.726 (Fixed)	\\
$x_{1,R}$, $x_{2,R}$	&	0.633, 0.633 (Fixed)	\\
$<r_{1}>$	&	0.487$\pm$0.001	\\
$<r_{2}>$	&	0.309$\pm$0.001	\\
$Co-Lat_{Spot I}$ ($^\circ$) =	&	60 (Fixed)	\\
$Long_{Spot I}$ ($^\circ$) =	&	210 (Fixed)	\\
$R_{Spot I}$ ($^\circ$) =	&	45 (Fixed)	\\
$T_{f Spot I}$ =	&	0.96 (Fixed)	\\
$Co-Lat_{Spot II}$ ($^\circ$) =	&	60 (Fixed)	\\
$Long_{Spot II}$ ($^\circ$) =	&	170 (Fixed)	\\
$R_{Spot II}$ ($^\circ$) =	&	45 (Fixed)	\\
$T_{f Spot II}$ =	&	0.96 (Fixed)	\\
\hline
\end{tabular} 
\end{table*}

\end{document}